\documentclass[a4paper, amsfonts, amssymb, amsmath, reprint, showkeys, nofootinbib, twoside]{revtex4-1}
\usepackage[english]{babel}
\usepackage[utf8]{inputenc}
\usepackage[colorinlistoftodos, color=green!40, prependcaption]{todonotes}
\usepackage{amsthm}
\usepackage{mathtools}
\usepackage{physics}
\usepackage{xcolor}
\usepackage{graphicx}
\usepackage[left=23mm,right=13mm,top=35mm,columnsep=15pt]{geometry} 
\usepackage{adjustbox}
\usepackage{placeins}
\usepackage[T1]{fontenc}
\usepackage{lipsum}
\usepackage{csquotes}
\usepackage[pdftex, pdftitle={Article}, pdfauthor={Author}]{hyperref} % For hyperlinks in the PDF
\DeclareUnicodeCharacter{2212}{-}
\begin{document}
%\title{Room temperature infrared photodetector based on electronic state engineered quantum dots}
%\title{Room temperature infrared photodetector operation by resonant state absorption in quantum dots}
%\title{Room temperature infrared detector operation by resonant state absorption in quantum dots}
%\title{Room temperature infrared detection with high detectivity by resonant state absorption in quantum dots}
\title{Conduction band resonant states absorption for quantum dot infrared detectors operating at room temperature}

\author{Stefano Vichi$^{1,*}$, Shigeo Asahi$^{2}$, Sergio Bietti$^{1}$, Artur Tuktamyshev$^{1}$, Alexey Fedorov$^{3}$, Takashi Kita$^{2}$, and Stefano Sanguinetti$^{1}$}
%    \email[Correspondence email address: ]{stefano.vichi@unimib.it}% Your name
%    \affiliation{University of Milano-Bicocca, Department of Materials Science, Via Cozzi 55,Milano, Italy}
%\author{Shigeo Asahi}
%    \affiliation{Department of Electrical and Electronic Engineering, Kobe University, 1-1 Rokkodai, Nada, Kobe 657-8501, Japan}
%\author{Sergio Bietti}
%    \affiliation{University of Milano-Bicocca, Department of Materials Science, Via Cozzi 55,Milano, Italy}
%\author{Artur Tuktamyshev}
%    \affiliation{University of Milano-Bicocca, Department of Materials Science, Via Cozzi 55,Milano, Italy}
%\author{Alexey Fedorov}
%    \affiliation{CNR–IFN, via Anzani 42, 22100 Como, Italy}
%\author{Takashi Kita}
%    \affiliation{Department of Electrical and Electronic Engineering, Kobe University, 1-1 Rokkodai, Nada, Kobe 657-8501, Japan}
%\author{Stefano Sanguinetti}
%    \affiliation{University of Milano-Bicocca, Department of Materials Science, Via Cozzi 55,Milano, Italy}

% Affiliations / Addresses (Add [1] after \address if there is only one affiliation.)
\address{%
1 \quad L-NESS and University of Milano-Bicocca, Department of Materials Science, Via Cozzi 55,Milano, Italy\\
2 \quad Department of Electrical and Electronic Engineering, Kobe University, 1-1 Rokkodai, Nada, Kobe 657-8501, Japan\\
3 \quad L-NESS and CNR–IFN, via Anzani 42, 22100 Como, Italy\\
* \quad corresponding author email: stefano.vichi@unimib.it
}

\date{\today} % Leave empty to omit a date

% AFM sample: C190124B
% IR sample: C201118

\begin{abstract}

Long Wavelenght infrared devices, despite growing interest due to a wide range of applications in commercial, public, and academic sectors, are still struggling to achieve significant improvements over well-established technologies like HgCdTe detectors. Devices based on quantum nanostructures remain non competitive due to unresolved drawbacks, the most significant being the need to cool down to liquid nitrogen temperatures to improve the signal-to-noise ratio.
In this work, we demonstrate an innovative solution to surpass the current generation of quantum dot-based detectors by exploiting the absorption from quantum dot localized states to  resonant states in the continuum, that is states in the semiconductor conduction band with an enhanced probability density in the quantum dot region. This unprecedented approach takes advantage of the unique properties of such states to massively enhance carrier extraction, allowing to overcome one of the most crucial drawbacks of quantum dot-based infrared detectors. This innovative solution is discussed here from both theoretical and experimental perspectives. The measured room temperature operation with high detectivity demonstrates that exploiting resonant states absorption in quantum dots offers the long-sought solution for the next generation of infrared photodetectors.

\end{abstract}

\keywords{room temperature, quantum engineering, quantum dots}

\maketitle

\section{Introduction} \label{sec:Introduction}

The advancement of long wavelength infrared (LWIR) detectors for imaging relies heavily on the realization of an improved generation of sensors. These innovative sensors aim to offer photodetectors with improved features, including a higher pixel count, high frame rates, superior thermal resolution, and the ability to operate at room temperature, thus having a groundbreaking impact on all domains, such as earth remote observation \cite{sobrino_minimum_2014,cao_review_2019}, automotive\cite{zhang_perception_2023}, imaging\cite{rogalski_optical_2004} and military defense. 

Currently, four primary semiconductor based photodetector technologies are being developed in the LWIR spectrum: HgCdTe photodiodes, quantum well infrared photodetectors (QWIP), quantum dot infrared photodetectors (QDIP), and antimonide-based type II superlattice photodiodes. Among these, bulk HgCdTe stands out as the most mature technology, offering superior performances and a broad, adjustable detection spectrum. On the other hand, the latter three technologies leverage quantum heterostructures based on III-V semiconductor materials, facilitating their integration with existing technology platforms\cite{rogalski_third-generation_2009}. Demonstrations of monolithic silicon integration for III-arsenic (As) and III-antimony (Sb) materials, fundamental for the implementation of these three technologies into actual devices, have been successful\cite{ren_recent_2019,kim_inasgaas_2017,bietti_fabrication_2009,cavigli_high_2012,bietti_monolithic_2013,ballabio_gaas_2019}. Among all, QDIP are gaining attention as a promising alternative, attributed to their precise control over transition energies due to three-dimensional carrier confinement and the lack of restrictive selection rules, which currently limit the broader adoption of QWIP technology. 

On the paper, QDIPs represent a significant advancement in the field of LWIR detection, offering a range of attractive capabilities: i)  high sensitivity, capable of detecting low levels of infrared radiation, ii)   engineered flexibility which allow to detect infrared radiation across a wide range of wavelengths by adjusting the size and composition of the Qds, iii)  low dark current and low noise characteristics, resulting in improved signal-to-noise ratios compared to other infrared detectors, iv)  can be fabricated using semiconductor well established processing techniques. This makes them suitable for various applications, including night vision, thermal imaging, and environmental monitoring. \cite{zeng_recent_2023}.  Despite their promising features, QDIPs also present several drawbacks that currently limit their performance and practical application.
One of the primary challenges associated with QDIP technology is the relatively low operating temperature. Currently, QDIPs often require cooling to very low temperatures to achieve optimal performance, which increases the complexity and cost of the detection systems. This cooling requirement is due to the thermal activation of carriers in the quantum dots, which can lead to increased noise and decreased detector sensitivity at higher temperatures.
Another limitation is the dark current, which is the current that flows through the photodetector even in the absence of light. In QDIPs, the presence of defects and the complex structure of the quantum dots can contribute to a higher dark current, thereby reducing the signal-to-noise ratio and impairing the detector's overall performance.
Additionally, the quantum efficiency, which is a measure of how effectively the photodetector converts incoming photons into electrical signals, can be lower in QDIPs compared to other infrared detection technologies. This is partly due to the inherent properties of quantum dots, due to the relatively low density of absorbers (the QDs) and to issues related to carrier escape and recombination within the dots.
While QDIPs offer the potential for high-resolution and tunable infrared detection, these drawbacks highlight the need for ongoing research and development to overcome the challenges and fully realize the technology's capabilities in practical applications.

With this work we are opening a new path for QDIP detection, which exploits the unique properties of resonant states absorption in quantum dots. This approach, which was never explored, allowed us to demonstrate a photodetector with a high room temperature detectivity in the LWIR spectral region. Our detector design specifically incorporates a hybrid structure comprising quantum dots and a quantum well, both made of the same material. This mixed structure results in hybrid QD-QW states, which can be engineered to enhance the extraction efficiency of photoexcited carriers while maintaining a significant wavefunction overlap with the QD ground state. Besides, this design provides the necessary control over quantum states and their carrier occupation to be able to selectively enhance or suppress electron thermalization channels, allowing to engineere carrier dynamics and the detection process in an unprecedented way. Here, we present the design principles, fabrication techniques, and experimental results of our novel quantum-based LWIR detector.

%\section{Results and Discussion} \label{sec:Results}

\section{Nanostructure design} \label{sec:Nanostructure_design}

Improving the current generation by LWIR at room temperature of QDIPs requires the a deep analysis of the design of the processes involved and of their weak points with the final goal of overcoming their limitations or limiting their impact on detection performances. 

\begin{figure*}
    \centering
    \includegraphics[width=\linewidth]{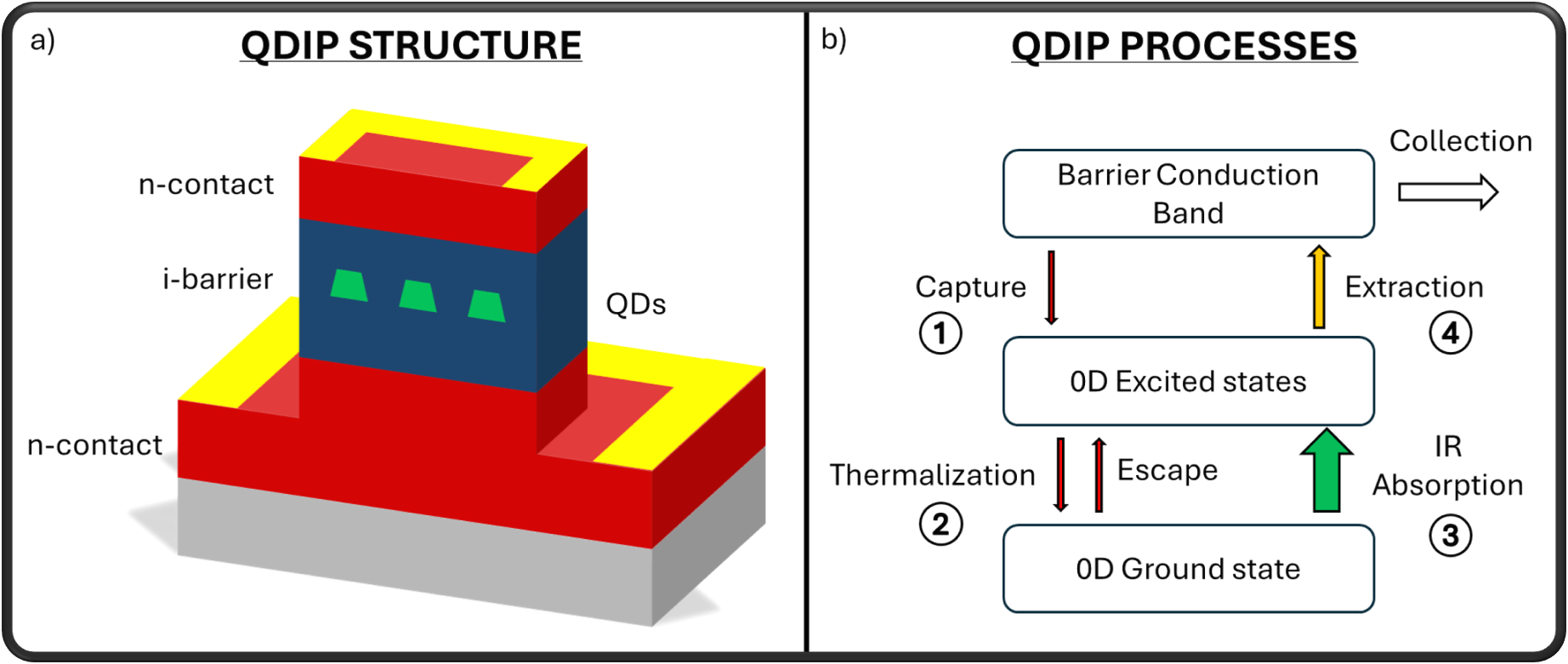}
    \caption{a) Typical QDIP device structure, made of an n-i-n heterojunction. One or more QD layers are embedded in the intrinsic barrier region with a larger band gap. b) Schematics of the electron dynamics processes in the QDIP. The numeric labels indicate the order of the detection steps. Each process is represented by and arrow which indicates the transition direction, with its color (from red to green) and size (from small to large) indicating an increasing transition probability.}
    \label{fig:qdip}
\end{figure*}

The typical QDIP device structure is illustrated in Figure \ref{fig:qdip} a. The QDIP is a photoresistive detector, where LWIR light is detected through an increase in device current due to photon absorption. This absorption causes transitions between the ground and excited states of electrons confined within the quantum dots (QDs). The standard electronic configuration is an n-i-n heterostructure. The active zone, embedded in the intrinsic region of the QDIP, consists of multiple layers of self-assembled QDs within a barrier layer with a larger bandgap, ensuring carrier confinement in three dimensions. The processes involved in QD intraband detection are schematized in Figure \ref{fig:qdip} b, providing a visual representation of the electron dynamics in the QDIP. When LWIR radiation strikes the QDIP, photons with energy matching the energy difference between the QDs confined states are absorbed. This absorption excites the electrons present in the QDs, which are introduced by intentional doping or are captured, and subsequently thermalized, from the barrier conduction band, from a lower energy state (often the ground state) to a higher energy state (an excited state) within the QDs. The specific energy levels of the quantum dots can be engineered to target different LWIR wavelengths, making QDIPs versatile for various IR detection applications. Once the electrons are excited to higher energy states, the electrons are promoted to the conduction band of the surrounding semiconductor material, by tunneling or thermionic processes, and are then collected by electrodes, creating a measurable photocurrent. Each process is indicated, in Figure \ref{fig:qdip} b, by an arrow, with its color (from red to green) and size (from small to large) indicating an increasing probability of the transition respectively. 

Let us begin by analyzing the critical process of carrier dynamics as depicted in Figure \ref{fig:qdip} b. The QDIP signal is generated through a series of processes: photon absorption, extraction of the photoexcited electron from the quantum dot (QD) into the conduction band, and finally, collection at the device electrodes. Competing with photon absorption is the thermally promoted carrier escape process, which is typically the dominant mechanism at room temperature in standard QDIP designs and adds up to the dark current generated by other areas of the device. For the QD to remain active after the initial photon absorption, the QD ground state must be refilled by electrons from the barrier conduction band through capture and thermalization processes. The efficiency of this combined process contributes to both photocurrent intensity and frame rate. In this regard, the small carrier capture cross section of QDs \cite{magnusdottir_one-_2002,muller_ultrafast_2003,ryzhii_theory_1996}, caused by their limited size, is certainly a week point of QDIPs. The color code in Figure \ref{fig:qdip} b, indicates the limiting processes (red color) in actual QDIP quantum device design.  

\begin{figure*}
    \centering
    \includegraphics[width=\linewidth]{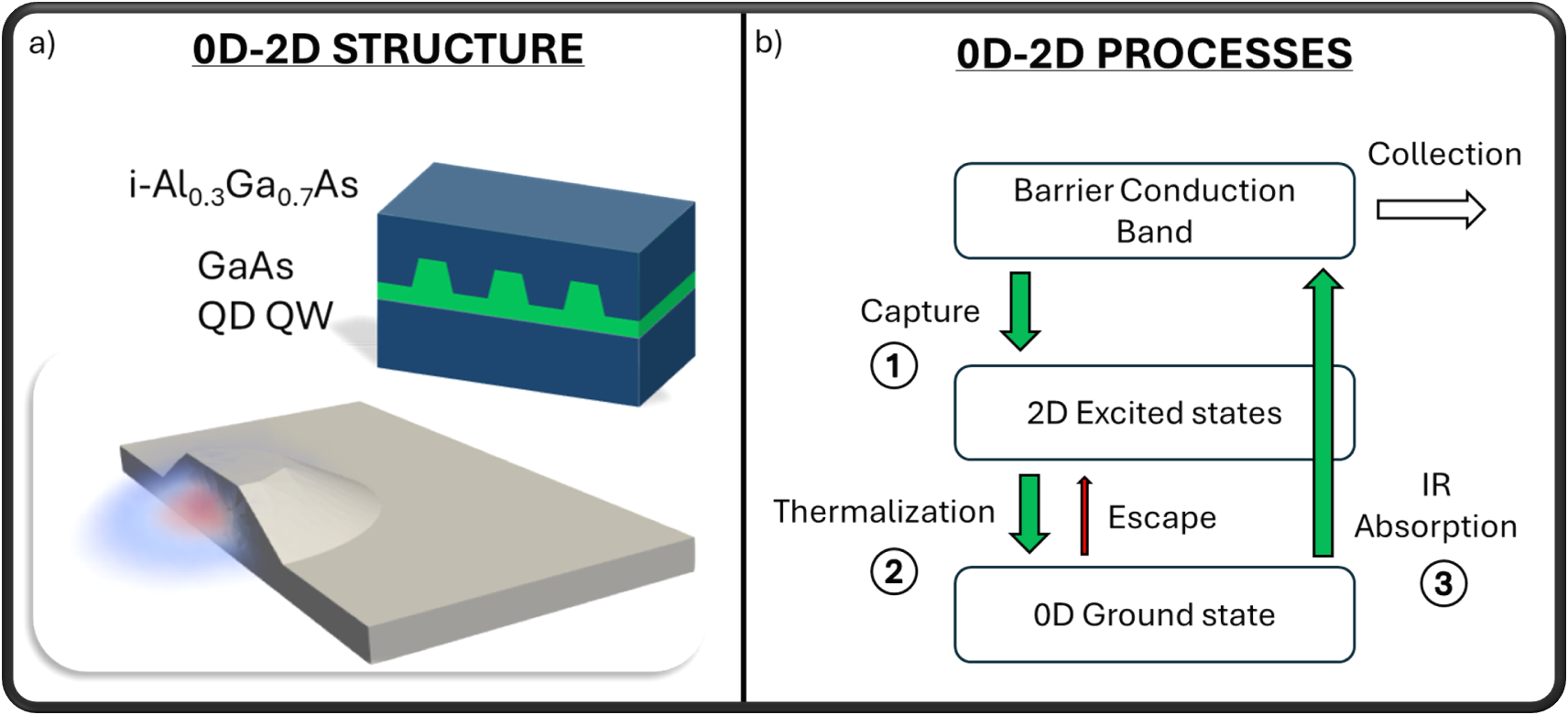}
    \caption{a) Representation of the 0D-2D nanostructure, with the drawing of the 0D ground state b) Schematics of the electron dynamics processes in the 0D-2D structure. The numeric labels indicate the order of the detection steps. Each process is represented by and arrow which indicates the transition direction, with its color (from red to green) and size (from small to large) indicating an increasing transition probability.}
    \label{fig:0d2d}
\end{figure*}

We begin our nanostructure design process by identifying corrective measures to mitigate the processes that limit QDIP performance at room temperature. The process of QD refilling, which involves the combination of carrier capture and thermalization to the ground state, has been identified in the literature as a limiting factor for QDIP sensitivity \cite{ryzhii_detectivity_2001}. In fact, if the electrons that leave the QD due to photon absorption or thermal excitation are not immediately replaced by electrons captured from the conduction band, the QD remains inactive, preventing further absorption.

In the literature, the capture probability drawback has been addressed in two ways: by increasing the number of free carriers in the barrier which can be captured (e.g. by using a pumping laser \cite{hoglund_optical_2009,ramiro_optically_2015,vichi_optically_2022}) or by increasing the capture probability of the QDs by embedding them in a QW of a material with a gap intermediate between the QDs and the barrier \cite{wolde_noise_2017}. The latter strategy, named dot-in-a-well (DWELL) device structure \cite{krishna_quantum_2005, barve_barrier_2012}, favours the carrier injection into the QDs by their spatial confinement in the QW. Still carrier scattering at the interface and strain induced potential barriers present in the DWELL design hinders a fast and efficient carrier capture in the QDs.

Here, the use of droplet epitaxy for QD fabrication - a self-assembling technique in an MBE environment based on the sequential deposition of group III and group V elements that does not rely on strain to promote QD formation \cite{koguchi_new_1991,kim_droplet_2018} — allows us to overcome these drawbacks. It is also extremely flexible and powerful when it comes to quantum state engineering, as it allows to obtain QDs with a narrow size distribution \cite{basso_basset_spectral_2019}, with an excellent control on their shape and size \cite{bietti_crystallization_2013,sablon_configuration_2008} and without restrictions on surface orientation \cite{heyn_droplet_2009,tuktamyshev_nucleation_2021,barbiero_exciton_2022}. This technique enables the combination of a quantum well (QW) with quantum dots (QDs) made of the same material, as depicted in figure \ref{fig:0d2d}  \cite{sanguinetti_modified_2003}. In particular the barrier material is an AlGaAs ternary alloy and the QW/QD nanostructure is made by GaAs. The resulting system is a thickness-modulated QW, with the 3D modulation controlled by the shape and size of the QDs.
This thickness-modulated QW possesses unique and complex electronic states that differ significantly from those in conventional QWs. The varying thickness creates regions where quantum confinement is significantly altered due to the reduced importance of the kinetic term in the electronic Hamiltonian in the thicker regions. As a result, electrons can become localized in regions with a reduced kinetic energy, leading to the presence of discrete QD-like energy levels at the bottom of the energy ladder. The higher-lying energy states remain QW-like.
This forms a hybrid QW/QD nanostructure in which the electronic states gradually transform from QD-like to QW-like states, with no distinct interfaces or potential barriers. A detailed calculation of such a hybrid structure is reported in Section \ref{sec:Quantum_properties}.

The use of a hybrid 0D-2D nanostructure not only improves the injection of carriers into the quantum dots (QDs) but also allows for extensive engineering flexibility. In fact, four independent structural parameters can be finely tuned in such systems, enabling precise control of the carrier processes within the nanostructure. These parameters are: (i) barrier height, (ii) quantum well (QW) thickness, (iii) QD lateral size, and (iv) QD height.

In a QDIP or a DWELL device, after the electrons are captured in a QD excited state, their thermalization to the QDs ground state is hindered by the phonon bottleneck effect \cite{steinhoff_combined_2013,scaccabarozzi_enhancing_2023,sanguinetti_dependence_2004,muller_ultrafast_2003}. The discrete density of states of QDs slows down this process as only transitions with an energy close to the one of optical phonons are allowed. This points out the inefficient thermalization process typical of QDs, which severely limits their ability to fill their ground state. It is possible to avoid such drawback by making the electronic state ladder with energy differencies between the states nearly matching the optical phonon energy (around 36 meV in GaAs) thus increasing the electron-phonon interaction \cite{steinhoff_combined_2013}. Typically, in InAs/InGaAs QD systems the electron-phonon interaction is strong enough to remove the phonon bottleneck as clearly demonstrated by the pinning of the quasi-Fermi energy in QD based intermediate band solar cells \cite{luque_experimental_2005}.

\begin{figure*}
    \centering
    \includegraphics[width=0.48\linewidth]{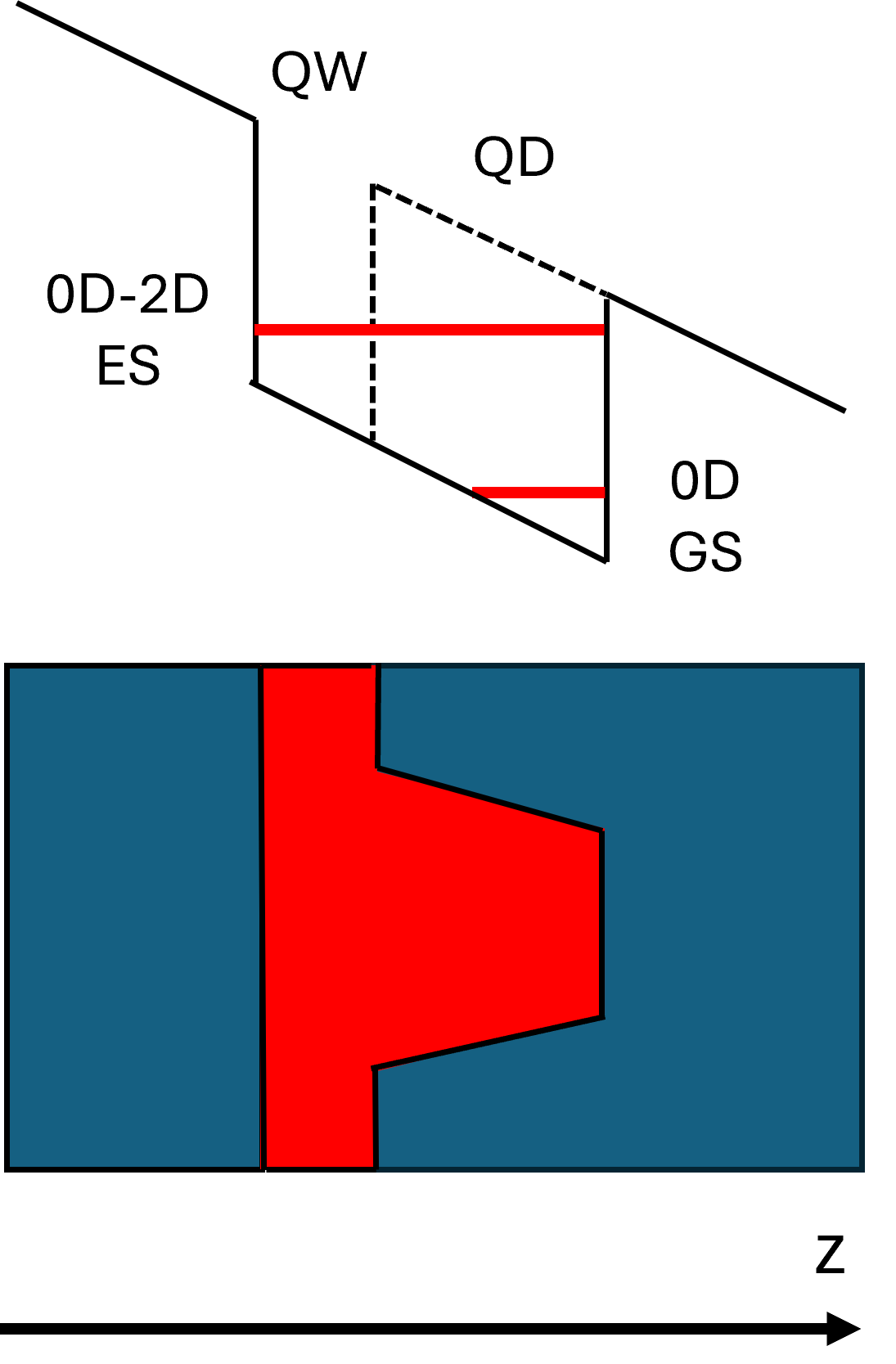}
    \includegraphics[width=0.48\linewidth]{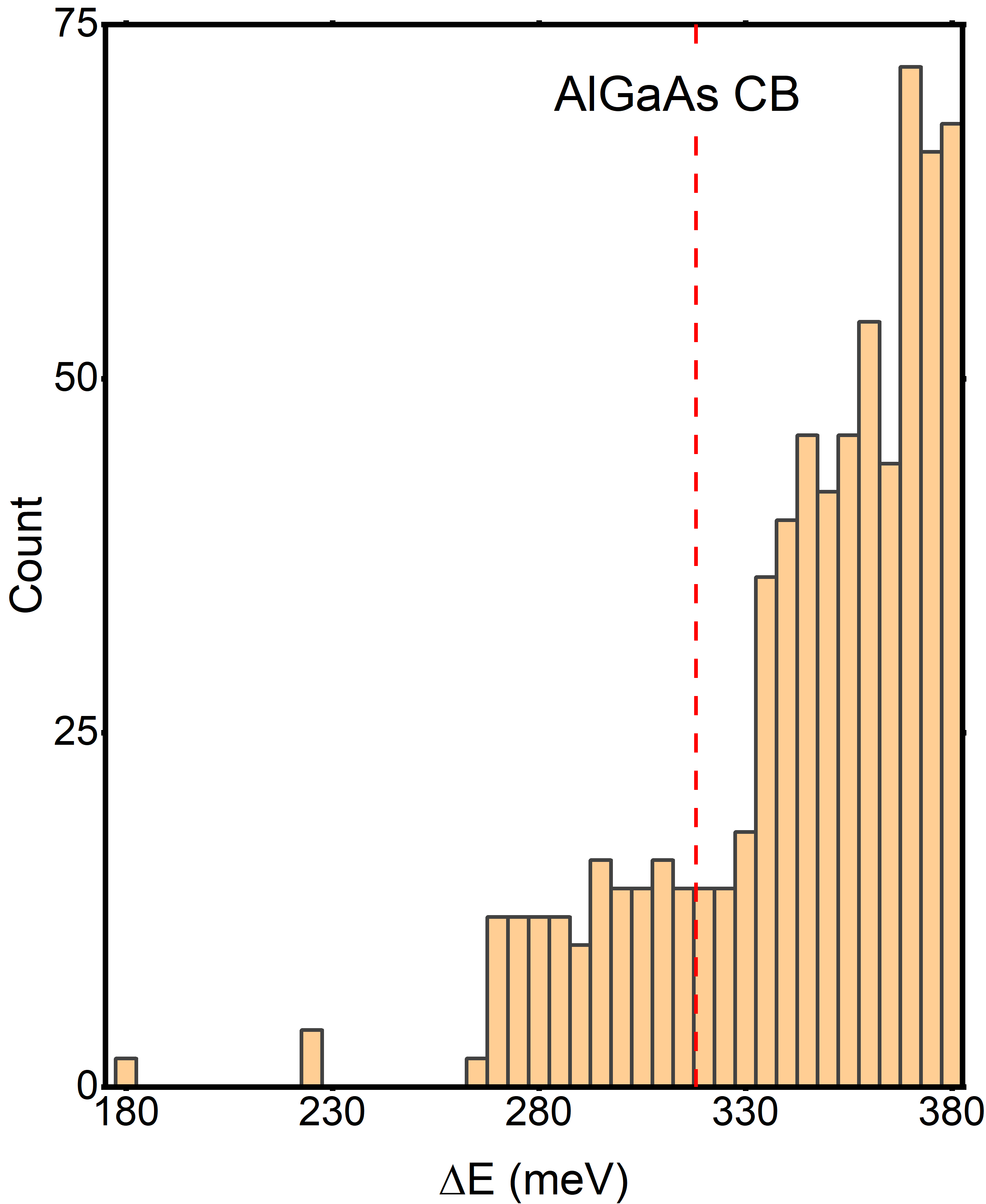}
    \caption{a) Schematized drawing of the ground state and excited confined energy levels in the 0D-2D nanostructure from where it can be intuitively understood the nature of the states. In particular, the ground state is zero dimensional while the excited states are mixed 0D-2D. b) Calculated electron density of states of the 0D-2D nanostructure, which demonstrates the claims on the confined states nature. The energies are reported with respect to the bottom of the GaAs conduction band. The dashed red line shows the energy of the Al$_{0.3}$Ga$_{0.7}$As CB.}
    \label{fig:nanostructure_scheme}
\end{figure*}

However, limiting the electron-phonon interaction helps suppress carrier thermal escape, which should significantly reduce the dark current, especially at higher temperatures. Therefore, the phonon bottleneck represents both an advantage and a disadvantage for QDIP performance. To benefit from it while avoiding its drawbacks, it is necessary to design the system to use different processes for carrier thermalization and thermal escape. The solution we propose again relies on the properties of the QW/QD heterostructure, which allows to engineer carrier dynamics in an unprecedented way with the objective of enabling alternative thermalization channels which are not relevant in QDIP and which do not involve the reduction of the electron-phonon interaction. In this regard, it is worth noting that the QW acts as a carrier reservoir for the QDs, thanks to the large density of state which can accommodate a large number of captured electrons. The absence of internal potential barriers compared to DWELL  design leads to the first significant difference with our engineered nanostructure. As can be seen from the schematic drawing of figure \ref{fig:nanostructure_scheme} a, thanks to our design the excited states have mixed 0D-2D nature, with a large wavefunction overlap with the 0D ground state. This feature, combined with the ability to accumulate a large number of carriers in a confined space, allows to increase significantly the probability of carrier-carrier scattering (i.e. Auger scattering), creating an alternative thermalization channel \cite{toda_efficient_1999,ferreira_carrier_1999}. Despite the possibility to have this process also in DWELL, the presence of an interface between QDs and QW combined with the two different confining potential results in a lower overlap of the states. As required, the presence of thermalization channels through carrier-carrier interaction does not affect the escape probability from the QDs ground state, maintaining the positive effects of phonon bottleneck.

Once the ground state of the QDIP is filled, electrons are able to absorb LWIR radiation by transitioning to an excited 0D confined state, with a probability determined by their optical matrix elements. This is another process in which QDIP shines, thanks to the large overlap between confined states. Once excited to a high-energy confined state, the electrons need to overcome a potential barrier before being collected at the contacts. This process can be achieved either by thermionic emission or by tunneling. Although these mechanisms are facilitated by temperature and bias applied to the device, their efficiency is limited and strongly depends on the position of the excited state with respect to the conduction band energy \cite{h_c_liu_dependence_1993}. Here our quantum design of the QDIP takes a different approach. As can be seen from the right panel of figure \ref{fig:0d2d}, we have designed the 0D-2D nanostructure in such a way that the transition occurs to a resonant state in the conduction band. This has a tremendous impact on the extraction probability of photoexcited carriers, as resonant states are conduction band states, thus eliminating completely carrier extraction issues.
The conduction band resonant states, which originate from the interference caused by the confining potential discontinuities on the continuum states, are well known in literature for 2D systems \cite{ikonic_bound-free_1989, hsu_bound_2016, azzam_photonic_2021, bogdanov_bound_2019, joseph_bound_2021} and are typically applied to resonant photonic devices such as lasers, sensors and filters \cite{fang_high-q_2021,romano_ultrasensitive_2020,yang_bending_2023}. In particular, due to the mixed nature of the QD-QW system devised in this study, here the resonant states in the in the conduction band are strongly mixed with the 0D confined states, thus being able to generates 0D-2D resonant states with a strong localization in the QD region resulting in large optical matrix elements for optical transitions with the QD-like ground state. This is the first time in which optical transitions to resonant states are used for IR detection in QDIPs.

\section{Quantum properties} \label{sec:Quantum_properties}
A more detailed insight on the quantum properties of the 0D-2D nanostructure can be understood by looking at the calculated conduction band density of states, based on the dimensions of our QDs, shown in figure \ref{fig:nanostructure_scheme}b. Here, the confined state energies are expressed with respect to the bottom of the conduction band. The low energy states are well separated while at higher energies the states form a continuous DOS with a step-like shape. These high-energy 2D states allow for a large carrier capture efficiency, act as a reservoir to fill the nanostructures ground state and allow to improve significantly the thermalization efficiency through carrier-carrier scattering \cite{toda_efficient_1999,ferreira_carrier_1999}.

\begin{figure*}
    \centering
    \includegraphics[width=\linewidth]{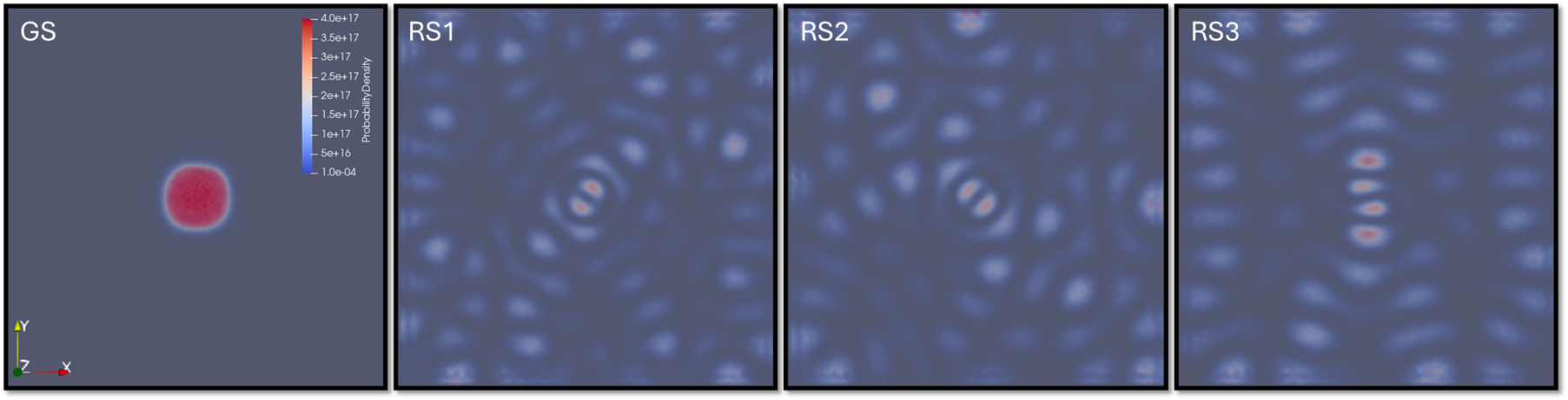}
    \includegraphics[width=\linewidth]{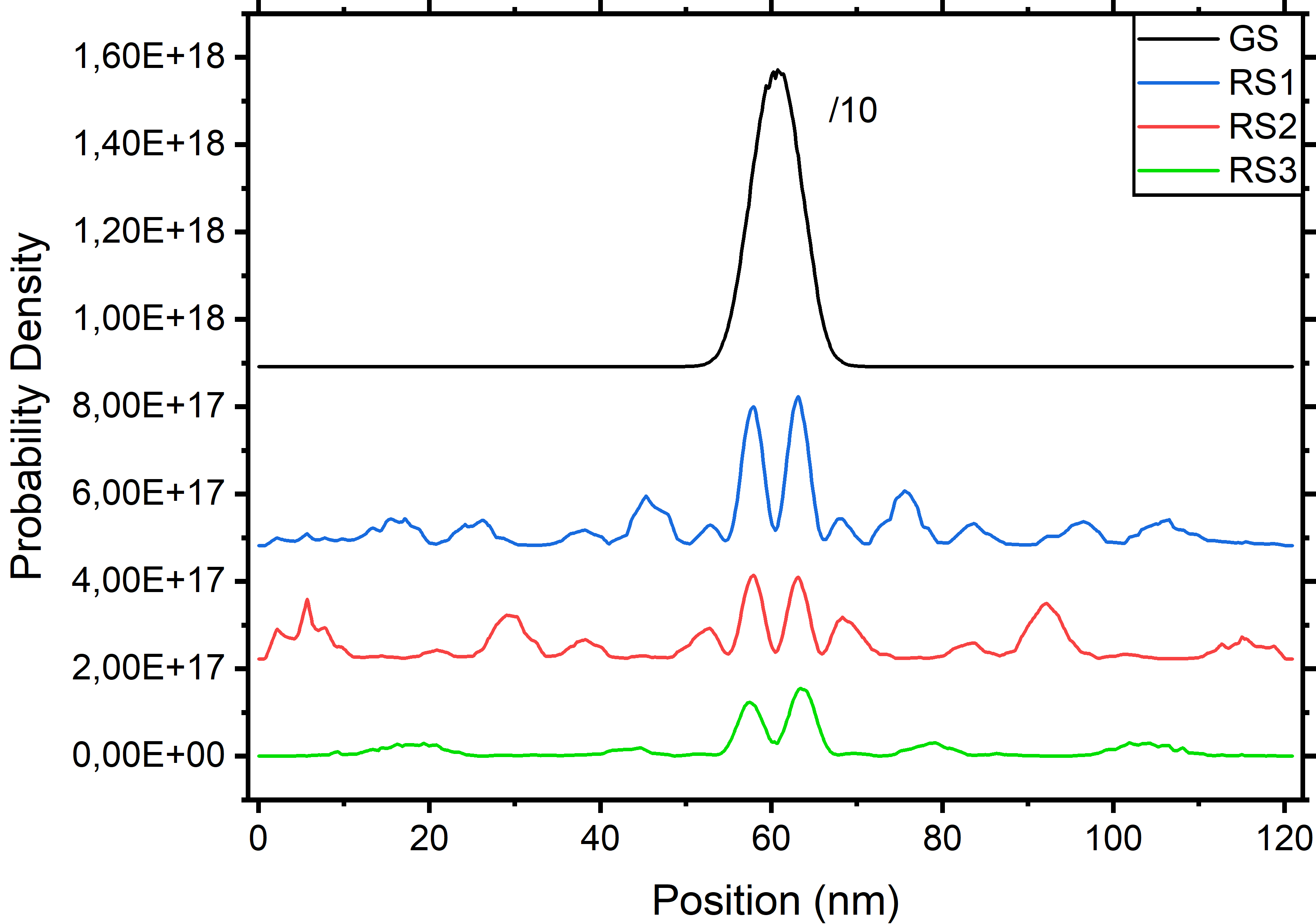}
    \caption{Electron wavefunction probability density of the ground state (GS) and the three excited resonant states with the largest optical matrix elements (RS1, RS2 and RS3 respectively). The transition energy with the ground state are 7.3$ \mu$m, 8 $\mu$m and 8.3 $\mu$m respectively. The bottom panel shows the line plot of the wavefunction probability densities along the diagonal.}
    \label{fig:q_states}
\end{figure*}

The resonant states above the CB edge play a significant role in the absorption of the 0D-2D hybrid nanostructure. Indeed, the numerical calculations performed on the basis of the designed nanostructures (see section \ref{sec:Device_characterization}) show the presence of a few resonant states with the largest optical matrix elements for the intraband transition with the ground state. Figure \ref{fig:q_states} shows the probability density function of the ground state and the three states with the largest optical matrix elements (from left to right respectively). As expected, the ground state is an s-like 0D state as typically found in QDs. On the other hand, all the excited states show a mixed 0D-2D component and a strong localization in the central region corresponding to the QD location (see the lower panel of figure \ref{fig:q_states}). These states have a transition energy with the ground state of 170 meV, 155 meV and 150 meV respectively, all of which are above the CB edge energy (138 meV). It is interesting to notice that none of these states has an s-like symmetry in the QD region but they all have a p-like symmetry, an expected  consequence of the intraband transition selection rules under dipole approximation \cite{luque_absorption_2012}.

\section{Device characterization} \label{sec:Device_characterization}

In order to obtain  absorption in the LWIR region, we designed the nanostructures with a QD height of 2.0 nm and a radius of 9.5 nm on top of a QW 1.1 nm thick. The material of the 0D-2D nanostructure was GaAs, while the barrier was made of Al$_{0.3}$Ga$_{0.7}$As.  

\begin{figure*}
    \centering
    \includegraphics[width=\linewidth]{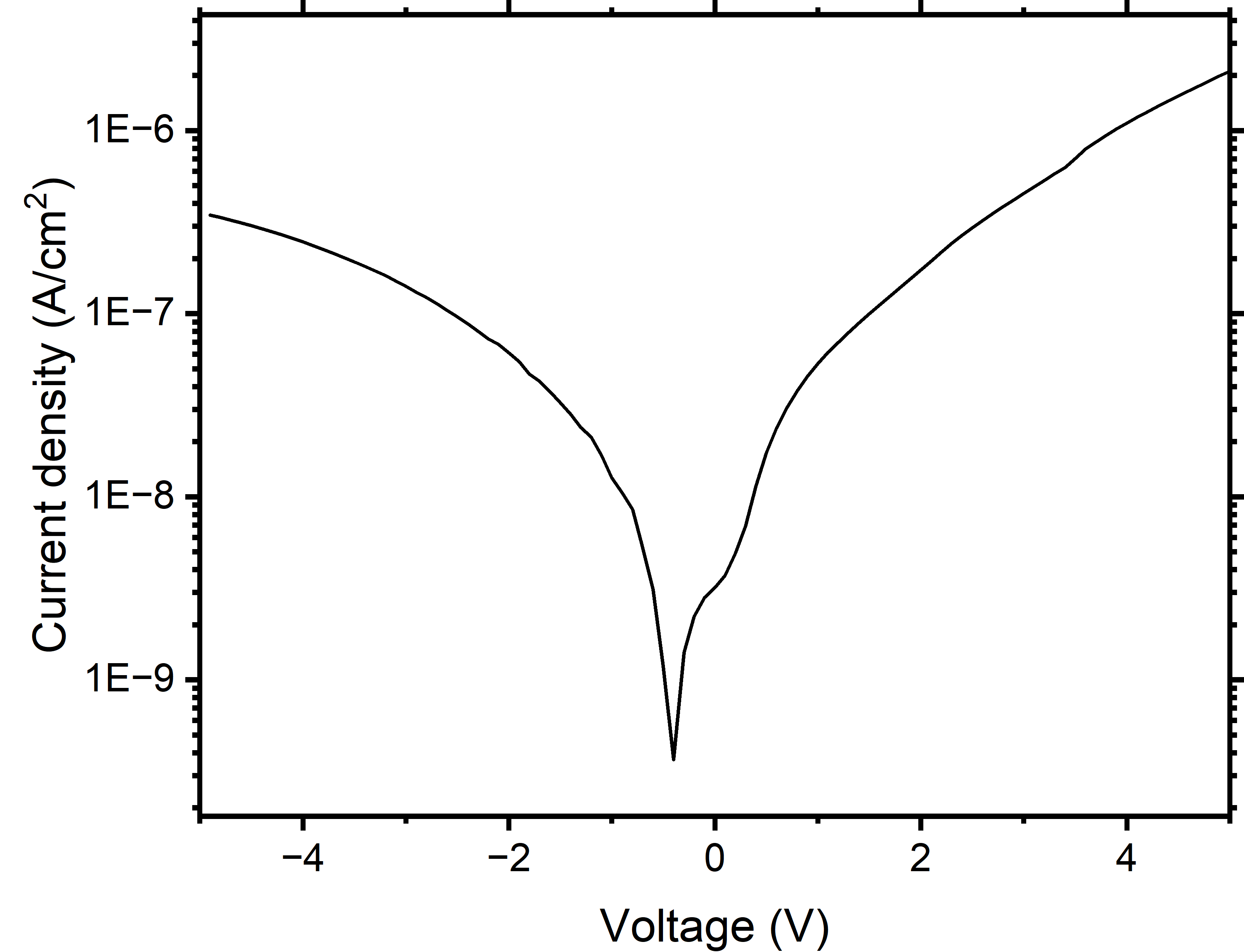}
    \caption{Room temperature I-V curve of the device.}
    \label{fig:IV}
\end{figure*}

The I-V curve of the device measured at room temperature is shown in figure \ref{fig:IV}. Remarkably, in our device the dark current is extremely low even compared to the low temperature measurements of QDIP \cite{wang_low_2001,jiang_06_2003,chakrabarti_high-performance_2005}, DWELL \cite{chen_demonstration_2018, ghadi_optimizing_2018} and QWIP \cite{billaha_influence_2016}. 
We attribute the slight shift of the zero-current point to the asymmetries in the stack of the grown structure which may lead to spurious capacitive effects. The asymmetry in the structure is also responsible for the different behavior at large voltage. 

\begin{figure*}
    \centering
    \includegraphics[width=.9\linewidth]{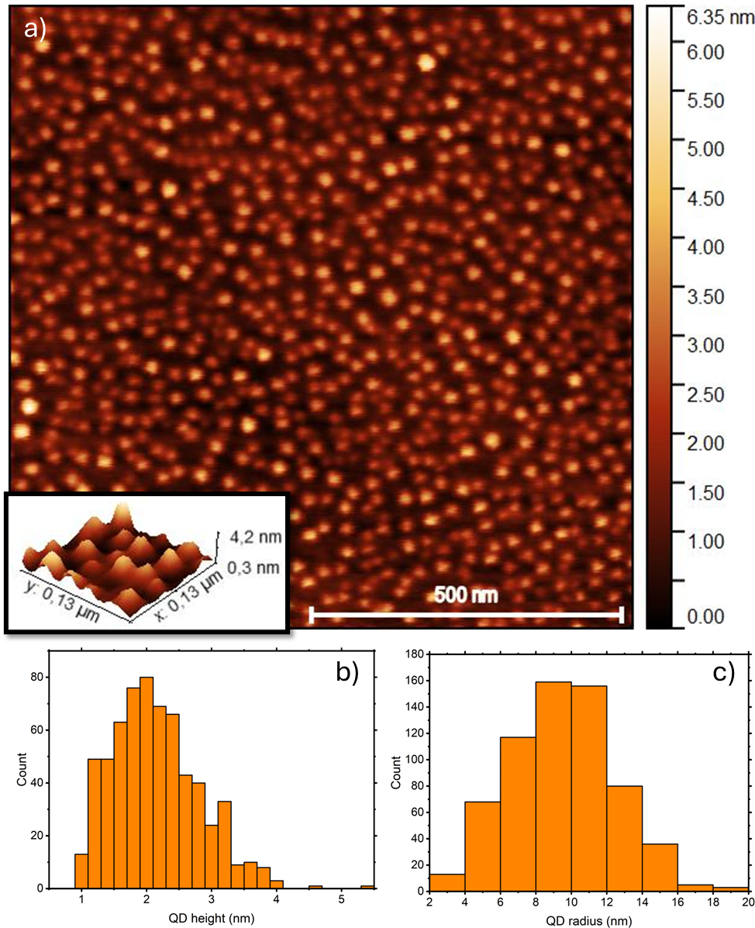}
    \caption{AFM image (a) and the obtained distribution of QD height (b) and radius (c). The mean QD height is 2.0$\pm$0.6 nm and the mean QD radius is 9.5$\pm$3.1 nm, with a dot density of $7\cdot10^{10} cm^{-2}$.}
    \label{fig:AFM}
\end{figure*}

Figure \ref{fig:AFM}a shows the AFM image of the uncapped QDs grown on a QW of 1.1 nm thickness. The mean height is 2.0$\pm$0.6 nm and the mean radius is 9.5$\pm$3.1 nm, as obtained by fitting with a Gaussian function the histogram data of figure \ref{fig:AFM} b) and c) respectively. The calculated QD density is $7\cdot10^{10} cm^{-2}$. The details of the AFM measurements are described in section \ref{ssec:AFM}. We attribute the relatively large size dispersion to the low substrate temperature and high Ga flux used for QDs formation.

\begin{figure*}
    \centering
    \includegraphics[width=\linewidth]{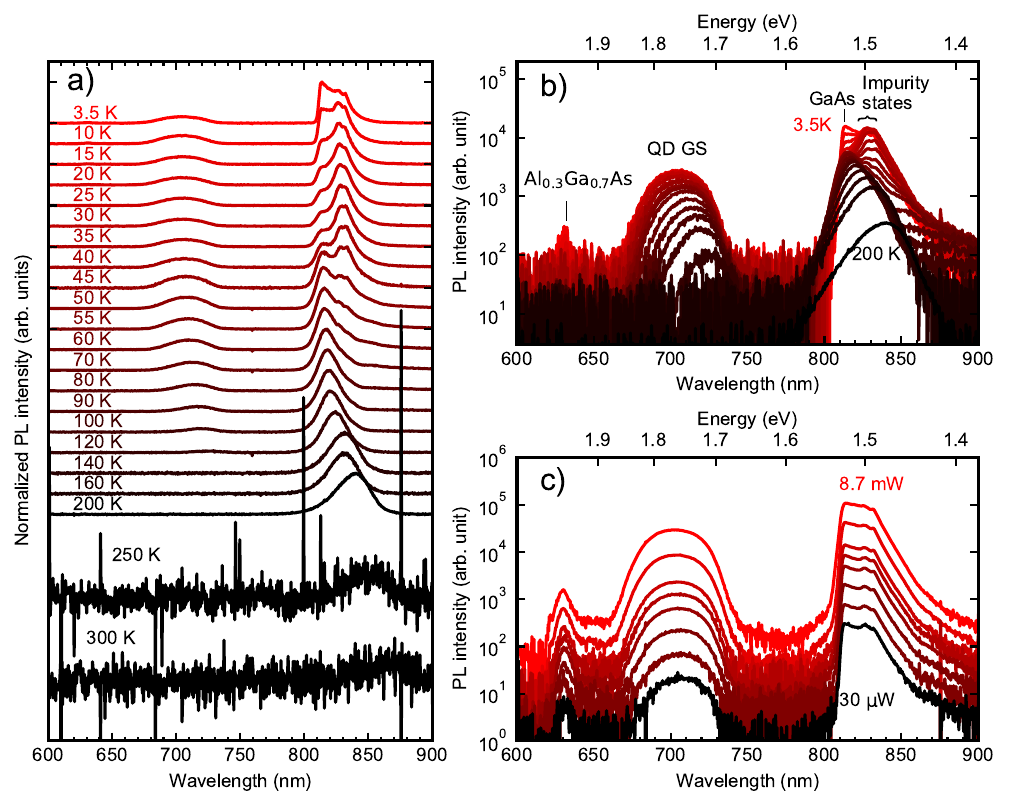}
    \caption{(a) Temperature dependence of normalized photoluminescence (PL) spectra. (b) Semi-logarithmic plot of figure \ref{fig:PL} a). (c) Intensity dependence of the PL at 4 K.}
    \label{fig:PL}
\end{figure*}

The optical properties of the QDs were investigated by photoluminescence (PL). Figure \ref{fig:PL} a displays the temperature dependence of the normalized PL spectrum, with the logarithmic plot of the PL spectrum presented in figure \ref{fig:PL} b. At 300 K, we observed an emission at the peak wavelength of 870 nm corresponding to 1.42 eV, which is attributed to the emission of GaAs. This emission intensifies and shifts to the shorter-wavelength side as the temperature decreases. When the temperature dropped below 45 K, the emission from GaAs was overwhelmed by a narrower-bandwidth emission, attributed to the exciton recombination in GaAs. The peak wavelength of the exciton emission was approximately 815 nm (1.52 eV) at 3.5 K. In addition, a distinct PL peak becomes evident with decreasing temperature at the wavelength of around 850 nm, appearing to be a combination of several peak PL emissions. These emissions originate from shallow impurity states due to carbon, Si, or other contaminants in GaAs \cite{pavesi_photoluminescence_1994}. Furthermore, an emission around 730 nm appears at 200 K, becoming significant and shifting to a lower wavelength side as the temperature decreases. The peak wavelength was approximately 710 nm at 3.5 K. This peak is attributed to the ground-state emission of QDs, in agreement with the calculated value of 704 nm obtained from the numerical simulations. Moreover, a weak PL emission was observed at the wavelength of 630 nm (1.97 eV), originating from the Al$_{0.3}$Ga$_{0.7}$As barrier layer.
Figure \ref{fig:PL} c shows the excitation intensity dependence of the PL spectrum at 4 K. The emissions from the QDs and the Al$_{0.3}$Ga$_{0.7}$As barrier layer become more evident with increasing intensity. Specifically, for the emission of the QDs, the lower-wavelength emission becomes more evident with increasing intensity. This feature indicates a state-filling of the QDs.

\begin{figure*}
    \centering
    \includegraphics[width=\linewidth]{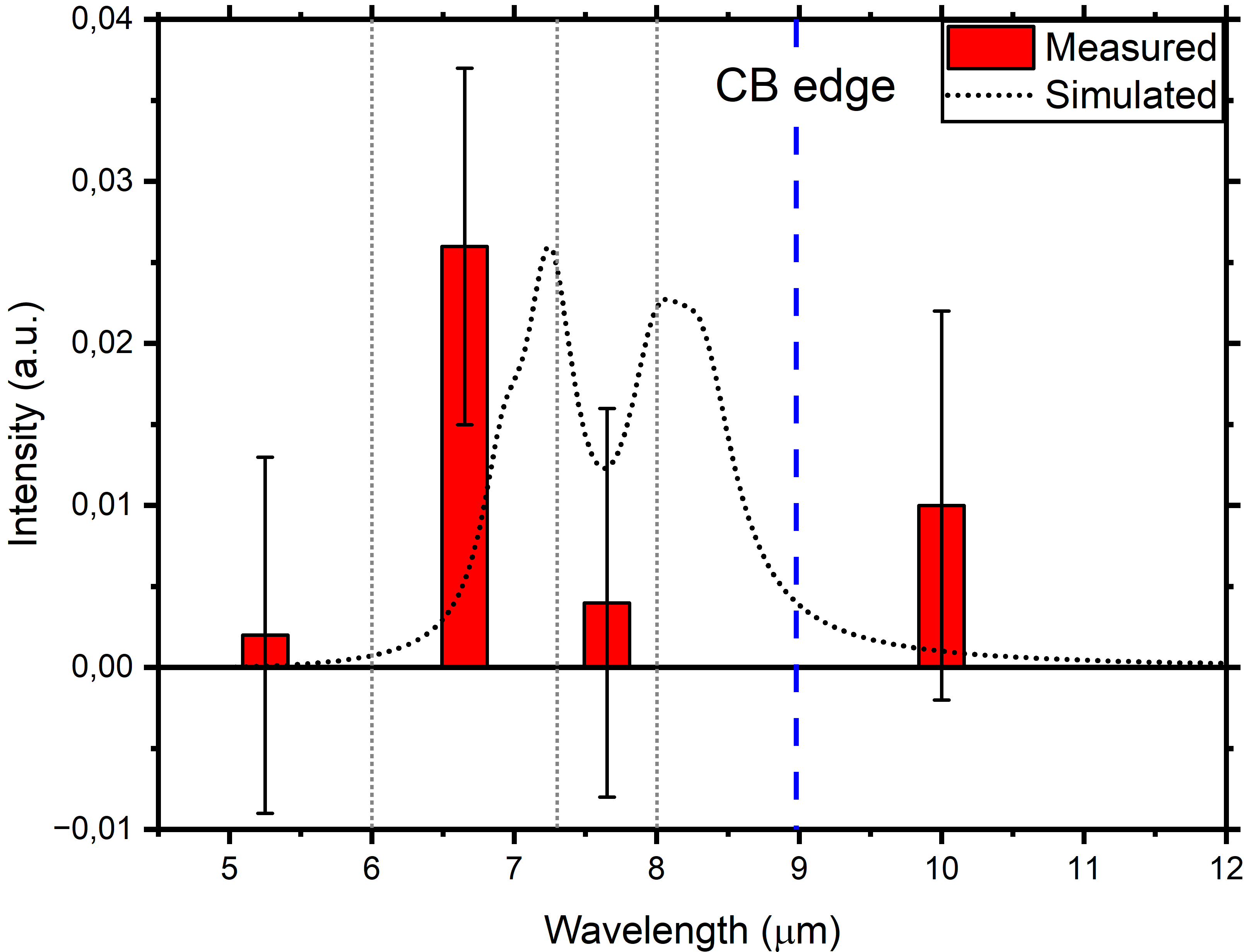}
    \caption{Comparison between the room temperature simulated IR absorption spectrum and the measured one. The dashed line indicates the offset between the electron ground state and the Al$_{0.3}$Ga$_{0.7}$As CB edge. }
    \label{fig:photocurrent}
\end{figure*}

Figure \ref{fig:photocurrent} shows the comparison between the simulated absorption spectrum (black line) and the measured spectrum at room temperature (red bars). As can be seen, there is a remarkable agreement between the simulations and the measurements. Indeed, the experimental results show that most of the absorption comes from shorter wavelengths than the transition from the GS to the CB edge, which confirms the theoretical prediction of an efficient absorption transition to resonant states in the conduction band. Remarkably, the main observed absorption peak is extremely narrow, lying in the 6 $\mu$m - 7.3 $\mu$m spectral window. It can be observed a second - less intense - absorption peak which lies between 8 $\mu$m and 12 $\mu$m. Also in this case our simulations confirm the resonant nature of the excited state responsible for this transition.
We measured a room temperature peak responsivity of $2 \cdot 10^{-4} A/W$ using the 6 $\mu$m longpass filter at a bias of -4 V.

\section{Discussion} \label{sec:Discussion}

The presence of resonant states in the continuum was already discussed in literature \cite{ikonic_bound-free_1989, hsu_bound_2016, azzam_photonic_2021, bogdanov_bound_2019, joseph_bound_2021} mostly related to QWs. Typically, the absorption spectrum involving these states is very broad due to the large amount of closely-spaced states \cite{ikonic_bound-free_1989}. On the other hand, our results show a narrow main absorption peak located in the 6 $\mu$m - 7.3 $\mu$m region. We attribute this feature to the combination of having a QD-like ground state and the presence of hybrid 0D-2D states at high energies, with a strong localization in the QD region. To explain this, we can describe each quantum state as a combination of an excited QD state and an excited QW state, with the quantum numbers of the two that are independent and determined by their total energy. Since the ground state of the system is a QD s-state, as can be seen from figure \ref{fig:q_states}, the absorption selection rules depend on the 0D component of each state, as this is the only one with non-zero matrix elements with the ground state. Therefore, since the selection rules for intraband transitions under dipole approximation require that the difference in the principal quantum number between the transitioning states must be an odd number, only the states with a 0D component which is p-like are allowed, as confirmed by the simulations of figure \ref{fig:q_states}. This causes a restriction in the number of states which can absorb light and therefore it limits the broadening of the absorption spectrum. 

With the knowledge of the optical and electrical performance is then possible to evaluate a fundamental figure of merit, the specific detectivity (D$^*$). The detectivity is a parameter defined to characterize the capability of detecting photons of a given detector. This capability is described by the ratio between the signal level and the noise level. D$^*$ can be calculated as follows 

\begin{equation}
    D^* = \frac{\sqrt{A \Delta f}}{NEP}
\end{equation}

where A is the detector area, $\Delta f$ is the bandwidth, and $NEP$ is the noise equivalent power. This is defined as the input power required of the optical signal for the signal-to-noise ratio to be unity (SNR = 1). The NEP in a photodetector can be expressed as the ratio between the mean square noise current at an input optical power for which SNR = 1 and its responsivity. In QD-based photodetectors operating at high temperature, it can be assumed that the major contribution to the noise current comes from the blackbody radiation in the background (i.e. thermal noise) \cite{martyniuk_quantum-dot_2008}. Therefore, the NEP is

\begin{equation}
    NEP = \frac{1}{\mathcal R}\frac{\sqrt{4 k_B T B}}{R}
\end{equation}

where $\mathcal R$ is the responsivity, $k_B$ is the Boltzmann constant, T is the temperature, B is the bandwidth, and R is the resistance. The photodetectors shows a D$^*$ of $9.8 \times 10^{9}$ cm Hz$^{1/2}$/W for an applied voltage of -4 V at T = 300 K. To our knowledge, this is the largest value reported for a QD-based LWIR photodetector operating at room temperature and only one order of magnitude lower than cooled devices \cite{rogalski_progress_2021,xue_high-operating-temperature_2023}. This result is even more remarkable considering the presence of a single layer of 0D-2D nanostructure in the device.

%Another great advantage of having a mixed QD-QW structure is the possibility to have a direct doping of the nanostructures by direct doping of the QW. In typical QDIPs the ground state of the QDs is filled by a delta-doping a few nanometers apart from the nanostructures. Donor ionization is required to actually fill the QDs and a local electric dipole forms between the QDs and the ionized donors, which may have a negative impact on the electrical properties of the device and on QD refilling \cite{}. On the other hand, direct doping of the QW avoids donor charging effects and improves the filling efficiency of the nanostructure ground state.

\section{Conclusions} \label{sec:Conclusions}

In this work we have presented an innovative approach to long wavelength IR detection based on resonant states absorption in QDs. By  deeply analizing QDIP  working principles we devised a quantum nanostructure whose electronic states have been engineered in order to exploit absorption transitions towards resonant states in the the continuum to massively enhance the extraction of photoexcited carriers. This innovative approach allowed us also to promote carrier capture and thermalization, while preventing thermionic emission from the QD ground states at room temperature. For this purpose we realized a hybrid 0D-2D nanostructure made of GaAs and embedded in an Al$_{0.3}$Ga$_{0.7}$As barrier layer with optimized QW thickness and QD shape and size. We realized such a device using the droplet epitaxy technique which allow for such a high degree of flexibility. 

The QDIPs realized using our design show a high room temperature detectivity of $9.8 \times 10^{9}$ cm Hz$^{1/2}$/W at -4 V peaked in the 6 $\mu$m - 7.3 $\mu$m range, thus demonstrating a remarkable agreement between the theoretical predictions and the experimental measurements. These figures show that the significant advantages in sensitivity and tunability of QDIPs can be brought to room temperature operation, thus realizing their full potential in various applications. Expanding into emerging markets and integrating with complementary technologies could present opportunities for the the growth and innovation in the LWIR sensor market in the future.

As a final remark, these results demonstrate the enormous potential of resonant states absorption in quantum dots, which is entirely unexplored up to date. With this work we are opening a new path for infrared photodetectors, which we believe will have a groundbreaking impact on many branches of optoelectronics.
\section{Methods} \label{sec:Methods}

\subsection{Growth} \label{ssec:Growth}
%sample  Matteo C 1 layer C190204; blank C190201; AFM C190124B
The sample was grown on a 2'' GaAs undoped wafer inside a Molecular Beam Epitaxy chamber provided with As valved cracked cell. 
The growth started with the deposition of 1 $\mu$m of Si doped GaAs as lower contact, followed by 350 nm of Al$_{0.3}$Ga$_{0.7}$As barrier layer deposited at 650°C and a 1.1 nm Si doped GaAs QW.
The quantum dots were fabricated on the top of the quantum well using droplet epitaxy. The substrate temperature was lowered to 150$^\circ$C and 3 MLs of Ga were deposited with a rate of 0.1 ML/s and a background pressure lower than $2\times10^{-9}$ torr. At the same temperature As valve was open to crystallize the Ga droplets with a beam equivalent pressure of $5\times10^{-5}$ torr.
The quantum dots were covered with 20 MLs of Al$_{0.3}$Ga$_{0.7}$As deposited by migration enhanced epitaxy at 400$^\circ$C and annealed for 5 minutes at this temperature to improve the crystalline quality.
The detector structure was completed with 350 nm of Al$_{0.3}$Ga$_{0.7}$As and 500 nm of Si doped GaAs as upper contact, deposited at 650°C.\newline

\subsection{Device fabrication} \label{ssec:Device fabrication}

The wafer was then processed in clean--room environment with photolithography and wet--etching process in order to create mesas and finally metal contacts were deposited in evaporation chamber.
Mesa etching was done by a wet approach, using an $H_{3}PO_{4}:H_{2}O_{2}:H_{2}O$ solution with 1:2.5:8 relative concentrations. Metal contacts were deposited by e-beam evaporation using a stack of 122 nm Cu and 78 nm Ge and annealed in nitrogen atmosphere at 420$^\circ$C C for 10 minutes, in order to get ohmic contacts. The mesa area was 0.09 $cm^{2}$.

\subsection{Photoabsorption} \label{ssec:Photoabsorption}

Photoabsorption measurements were performed on a dedicated system composed by an infrared source operating at 900$^\circ$C, a parabolic mirror and a plane--convex ZnSe lens to focus the infrared radiation, a chopper and a set of optical longpass filters at 4.5, 6 and 7.3 $\mu$m and a bandpass between 8 and 12 $\mu$m).
The sample was connected to a voltage generator, a transimpedance amplifier with a fixed gain of $10^{6} A/V$ and a lock--in amplifier.
For each filter we measured the output signal performing a sweep between -4 and 0 volts, obtaining an I-V characteristics from which we calculated the resistance of the detector under IR radiation. When using longpass filters, the resulting values include the contributions from longer wavelengths. Therefore, we determined the change in resistance in each spectral window by subtracting the results obtained with two subsequent filters. This value is inversely proportional to the photogenerated current.

\subsection{AFM} \label{ssec:AFM}

The sample morphology was analyzed at room temperature by the Veeco Innova atomic force microscope (AFM) in tapping mode using Nanosensors SSS-NCHR sharp silicon tips capable of a lateral resolution of about 2~nm.  The AFM images were then processed and analyzed with the software Gwyddion \cite{necas_gwyddion_2012}. The QDs (i.e. grains) were then identified by defining a height threshold in the image. Their equivalent radius was obtained by measuring the width of the grains at the intersection with the threshold. Their height was obtained by subtracting the median height of the background after subtracting the grains to the maximum value of each grain.

\subsection{Numerical simulations} \label{ssec:Numerical simulations}

The numerical simulations of the quantum properties of the nanostructures were performed using TiberCAD \cite{noauthor_see_nodate}. The calculations were performed using a finite elements method, modeling the nanostructures based on the geometrical parameters obtained from the analysis of AFM images (figure \ref{fig:AFM}). The QD shape was assumed to be a truncated cone, as described in a previous work of Bietti et al. \cite{bietti_precise_2015}, with an height of 2.0 nm and a major radius of 9.5 nm. The band structure was obtained by solving single-electron drift diffusion equations including Pikus-Bir strain corrections. The quantum states were then computed in the structure by using the k$\cdot$p model with an eight band envelope function approximation. Finally, the intraband transitions were studied by computing the optical matrix elements of the transitions in the dipole approximation, as defined in the equation:

\begin{equation}
    \abs{M}^2_{fi} = \abs{\bra{\Psi_f} \boldsymbol{P} \cdot \hat{e} \ket{\Psi_i}}^2
\end{equation}

where $\Psi_{f,i}$ are the envelope functions of the final and initial state respectively, $\boldsymbol{P} = -i \hbar \grad$ is the photon momentum operator and $\hat{e}$ is the polarization versor \cite{harrison_quantum_2016}. In our setup, where the IR light was incident perpendicularly on the surface of the detector, only the in-plane polarization components are taken into account.
Materials parameters for the simulations were taken from Vurgaftman et al.\cite{vurgaftman_band_2001}. 

%The calculation of the quantum states was made without taking into account the external bias applied to the structure. This choice was imposed by the excessive computational cost that this solution would have required. Indeed, the computation of the resonant states requires the calculation of a large number of continuum states at lower energies. In presence of an external potential that causes the bending of the electronic bands, the constraints of the simulation bounding box causes the appearance of several non-physical states that would further increase the computational cost.

\subsection{Photoluminescence} \label{ssec:Photoluminescence}

For the photoluminescence measurement a continuous-wave mode solid-state laser with a wavelength of 532 nm was used. For the detection, an Ocean Optics USB2000+ spectrometer with a focal length of 42 mm for input and 68 mm for output was used. The temperature was controlled by using a closed-cycle He cryostat.

\section{Acknowledgement}
We acknowledge support from MUSA – Multilayered Urban Sustainability Action – project (ECS 000037), funded by the European Union – NextGenerationEU, under the National Recovery and Resilience Plan (NRRP) Mission 4 Component 2 Investment Line 1.5: Strenghtening of research structures and creation of R\&D “innovation ecosystems”, set up of “territorial leaders in R\&D”

\section{Author contributions}

S.V. and S.B. performed the simulations. S.B., A.T. and A.F. carried out the growth and performed the AFM measurements. S.A. and T.K. performed PL measurements and analysis. S.B. performed the photoabsorption measurements. S.V., S.B. and S.S. have conceived the idea. All authors contributed to the data analysis and to the writing of the paper.

\section{Competing interests}
The authors declare no competing interests.

\section{Correspondence}
Correspondence and requests for materials should be addressed to
Stefano Vichi.

\section{Data availability}
The data that support the findings of this study are available from the corresponding author upon reasonable request.

\bibliography{Biblioteca}

\end{document}